# Automated Lesion Segmentation in Whole-Body FDG-PET/CT with Multi-modality Deep Neural Networks

Satoshi Kondo[1] and Satoshi Kasai[2]

[1] Muroran Institute of Technology, Hokkaido, Japan
[2] Niigata University of Health and Welfare, Niigata, Japan

**Abstract.** Recent progress in automated PET/CT lesion segmentation using deep learning methods has demonstrated the feasibility of this task. However, tumor lesion detection and segmentation in whole-body PET/CT is still a challenging task. To promote research on machine learning-based automated tumor lesion segmentation on whole-body FDG-PET/CT data, Automated Lesion Segmentation in Whole-Body FDG-PET/CT (autoPET) challenge is held, and a large, publicly available training dataset is provided. In this report, we present our solution to the autoPET challenge. We employ multi-modal residual U-Net with deep super vision. The experimental results for five preliminary test cases show that Dice score is $0.79 \pm 0.21$.

## 1    Introduction

Recent progress in automated PET/CT lesion segmentation using deep learning methods has demonstrated the feasibility of this task. However, tumor lesion detection and segmentation in whole-body PET/CT is still a challenging task. One bottleneck for progress in automated PET lesion segmentation is the limited availability of training data. To promote research on machine learning-based automated tumor lesion segmentation on whole-body FDG-PET/CT data, Automated Lesion Segmentation in Whole-Body FDG-PET/CT (autoPET) challenge is held, and a large, publicly available training dataset is provided.

In this report, we present our solution to the autoPET challenge. We employ residual U-Net with deep super vision with multi-modality fashion.

## 2    Proposed Method

The input data for lesion segmentation are whole-body PET/CT volumes, and two volumes, i.e., CT and SUV (Standardized Uptake Value) which is obtained from PET, are provided for each case. CT and PET volumes are acquired simultaneously on a single PET/CT scanner in one session, thus CT and PET (SUV) volumes are anatomically aligned up to minor shifts due to physiological motion.

We use 3D encoder-decoder networks for the segmentation task. Our base model is residual U-Net with deep super vision [2]. Input two volumes are resampled in [2 mm, 2 mm, 3 mm] for x, y and z direction, respectively, at first. CT and SUV volumes are normalized. The minimum and maximum values are -100 and 250, respectively, for CT volumes. And the minimum and maximum values are 0 and 15, respectively, for SUV volumes. In the training phase, we randomly sample 3D patches from the input volumes. The size of a 3D patch is 48 x 48 x 32 voxels. We sample 12 patches from each volume. When the volume includes lesions, the ratio of positive and negative patches in the sampling for one input volume is 3:1. We do not apply any augmentation. The patches of CT and SUV are concatenated to one patches as 2 channel patches, and then the concatenated patches are fed into the segmentation network.

The loss function is a weighted summation of Dice loss and cross entropy loss. The weights for Dice and cross entropy losses are 1 and 0.5, respectively. We also employ deep super vision for loss calculation. Intermediate outputs from several layers in the decoder of the model are up-sampled, loss value is calculated for each up-sampled output, and then the loss values are aggregated. The number of layers used in the deep super vision is two.

We train multiple models. Each model is trained independently using different combinations of training and validate datasets, and the inference results are obtained by ensemble of the outputs from the models. The final likelihood score is obtained by averaging the likelihood scores from the models. We use five models in our experiments.

## 3 Experiments

The dataset for training consists of 1014 studies of 900 patients acquired on a single site. The dataset for preliminary evaluation consists of 5 studies.

Our method is implemented by mainly using PyTorch [3], PyTorch Lightning and MONAI libraries. We use three Nvidia RTX3090 GPUs for training.

For the training of the segmentation model, the optimizer is Adam [4] and the learning rate changes with cosine annealing. The initial learning rate is 0.001. The number of epoch is 300. The model taking the lowest loss value for the validation dataset is selected as the final model.

We evaluated our method with the evaluation system provided by the organizers of the autoPET challenge. There are three evaluation metrics. The first one is foreground Dice score of segmented lesions. The second one is volume of false positive connected components that do not overlap with positives, i.e., false positive volume. The final one is volume of positive connected components in the ground truth that do not overlap with the estimated segmentation mask, i.e., false negative volume.

The results of our submission are Dice score is $0.79 \pm 0.21$, false positive volume is $0.29 \pm 0.66$, and false negative volume is $14.27 \pm 17.31$.

## 4 Conclusions

In this report, we presented our solution for the autoPET challenge. We employ multi-modal residual U-Net with deep supervision. The experimental results for five preliminary test cases show that Dice score is 0.79 ± 0.21.

## References

1. https://autopet.grand-challenge.org/
2. Futrega, M., Milesi, A., Marcinkiewicz, M., Ribalta, P.: Optimized U-Net for Brain Tumor Segmentation. arXiv preprint arXiv:2110.03352 (2021).
3. Paszke, A., Gross, S., Massa, F., Lerer, A., Bradbury, J., Chanan, G., ... & Chintala, S. (2019). Pytorch: An imperative style, high-performance deep learning library. Advances in neural information processing systems, 32.
4. Kingma, DP., Ba, J.: Adam: A method for stochastic optimization. In Proceedings of 3rd International Conference on Learning Representations (2015).